\documentclass[aps,prb,twocolumn,superscriptaddress,10pt,floatfix]{revtex4-2}

\usepackage{amsmath,amssymb}
\usepackage{graphicx}
\usepackage{braket}
\usepackage{xcolor}
\usepackage{adjustbox}

\usepackage{hyperref}
\definecolor{clink}{RGB}{0,70,255}
\hypersetup{colorlinks=true,linkcolor=clink,citecolor=clink,urlcolor=clink}

\begin{document}

\title{AKLT Hamiltonian from Hubbard tripods}

\author{Claire Benjamin}
\affiliation{Department of Physics and Astronomy, University of California, Irvine}
\author{Dániel Varjas}
\affiliation{Department of Theoretical Physics, Institute of Physics, Budapest University of Technology and Economics}
\affiliation{Institute for Theoretical Solid State Physics, IFW Dresden and Würzburg-Dresden Cluster of Excellence}
\author{Gábor Széchenyi}
\affiliation{Department of Materials Physics, ELTE Eötvös Loránd University}
\affiliation{HUN-REN Wigner Research Centre for Physics}
\author{Judit Romhányi}
\affiliation{Department of Physics and Astronomy, University of California, Irvine}
\author{László Oroszlány}
\affiliation{Department of Physics of Complex Systems, Eötvös Loránd University}
\affiliation{HUN-REN Wigner Research Centre for Physics}

\date{\today}

\begin{abstract}
We investigate how the spin-1 Affleck--Kennedy--Lieb--Tasaki (AKLT) Hamiltonian can emerge from a microscopic fermionic model based on half-filled Hubbard tripods. We first show that a single tripod hosts a robust threefold-degenerate low-energy manifold corresponding to an effective $S=1$ degree of freedom. This manifold prevails over a broad range of interactions and remains stable against moderate disorder. We then combine exact diagonalization with fourth-order quasi-degenerate perturbation theory to derive an effective bilinear--biquadratic spin model for a pair of coupled tripods and identify coupling regimes where the target ratio is approached. In particular, tuning leg-center hopping together with two symmetry-inequivalent leg-leg hoppings yields the characteristic singlet--triplet degeneracy associated with a biquadratic-to-bilinear ratio close to $1/3$. Extending the analysis to three tripods, we compare nonequivalent coupling geometries and find a strategy that suppresses unwanted longer-range and multispin terms while preserving the target nearest-neighbor couplings in the weak-coupling regime. These results establish a concrete bottom-up route from Hubbard clusters to valence-bond-solid spin physics in tunable quantum-dot arrays.
\end{abstract}

\maketitle

\section{Introduction}\label{sec:Intro}

Since Haldane's seminal work, antiferromagnetic integer-spin chains play a central role in condensed matter physics as they realize gapped quantum phases with nontrivial topology and fractionalized edge states \cite{haldane1983_prl,haldane1983_pla}. The AKLT construction provides an exactly solvable representative of the Haldane phase and therefore serves as a key microscopic reference point for symmetry-protected topological phases in spin systems \cite{AKLT}.

Beyond their significance in many-body physics, AKLT states are also directly relevant for quantum information processing. In measurement-based quantum computation (MBQC), one performs adaptive local measurements on a highly entangled resource state to implement a computation \cite{raussendorf2001}. In this context, the AKLT state provides a physically motivated resource candidate for understanding the role of restricted-resource states \cite{brennen2008_aklt}.

Realizing such effective quantum-spin models in controllable solid-state platforms is a major challenge. Semiconductor spin-qubit architectures and quantum-dot arrays provide a promising route due to rapidly improving tunability of local potentials, tunnel couplings, and geometry \cite{loss1998,hanson2007,john2024}. This motivates a bottom-up approach in which desired spin Hamiltonians are engineered from underlying fermionic models.

The single-orbital Hubbard model on coupled star-shaped clusters is a natural starting point for constructing effective quantum spins and their corresponding low-energy spin Hamiltonians. Lieb's theorem states that in bipartite lattices at half filling, the sublattice imbalance determines the degeneracy and spin of the ground-state manifold~\cite{lieb1989}. For star-like geometries, this mechanism enables robust emergent moments, as shown in early analytic explorations of the Hubbard-star \cite{vanDongen1991_single_star}, inspiring recent work on spin-1 chain physics from correlated-electron models \cite{Catarina_chain_PhysRevB.105.L081116,Abdelwahab_chain_PhysRevB.111.075129}. 

\begin{figure}[t]
  \centering
  \adjustbox{valign=c}{\includegraphics[width=0.2\columnwidth]{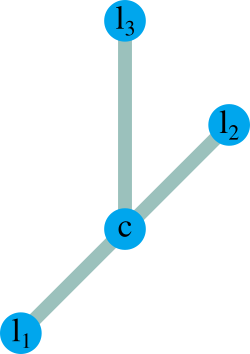}}
  \hspace{1em}
  \adjustbox{valign=c}{\includegraphics[width=0.7\columnwidth]{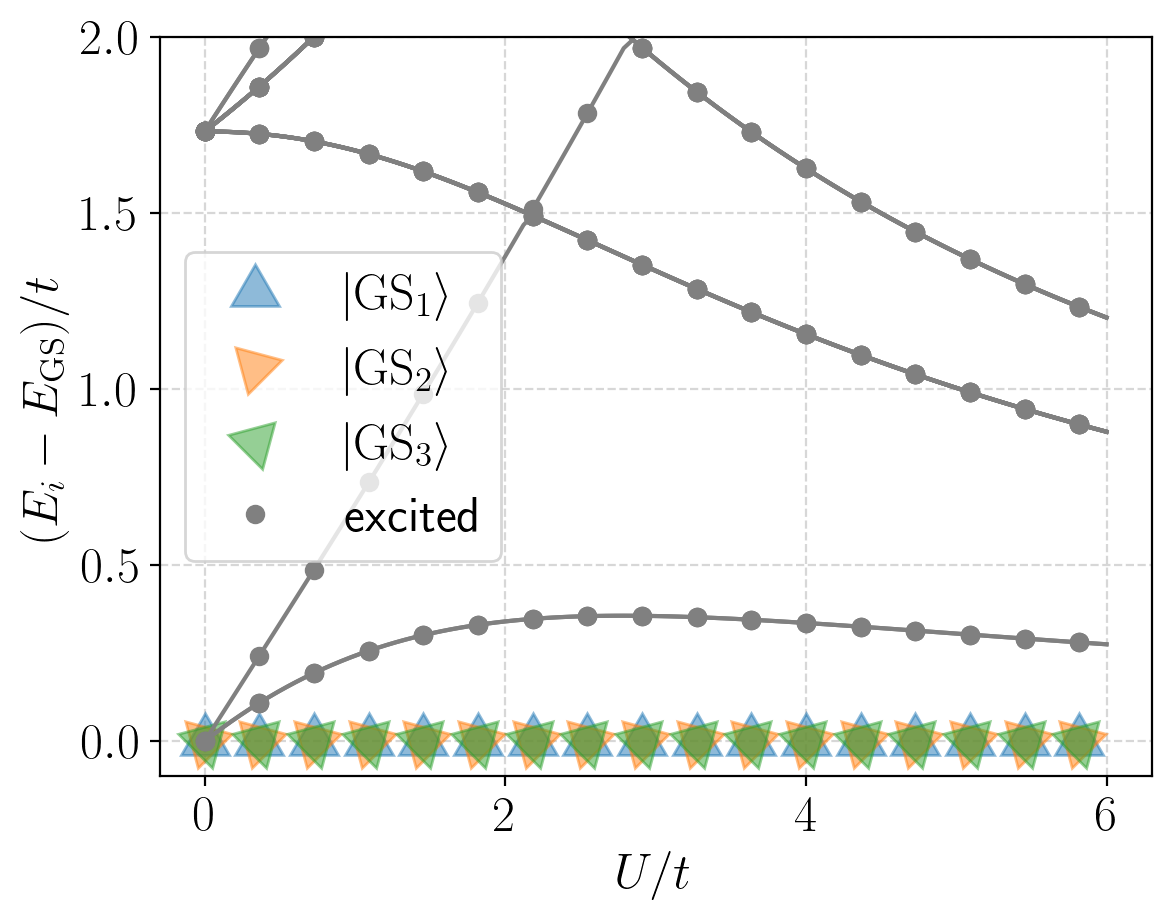}}
  \caption{The left panel depicts the schematics of a single Hubbard tripod with central site $c$ and legs $l_1$, $l_2$, $l_3$. The right panel shows the excitation spectrum of a single Hubbard tripod at half filling,  as a function of $U$.}
  \label{fig:tripod_spectrum}
\end{figure}

In this paper, we consider a Hubbard-star construction of quantum dots at half-filling dubbed tripods (see left panel of Fig.~\ref{fig:tripod_spectrum}), which provide the $S=1$ degrees of freedom as the building blocks of an effective quantum spin model, to achieve the AKLT Hamiltonian 
\begin{align}
    \mathcal{H}_{\rm AKLT}=J\sum_i \left [\mathbf{S}_i\cdot\mathbf{S}_{i+1}+\frac{1}{3}\left (\mathbf{S}_i\cdot\mathbf{S}_{i+1}\right)^2\right ].
    \label{eq:AKLT_chain}
\end{align}
Starting with the Hubbard tripods, we explore how they need to be assembled to produce an effective bilinear-biquadratic spin model corresponding to \eqref{eq:AKLT_chain}, with the desired ratio of the biquadratic and bilinear couplings, essential for realizing the desired resource state. 

\section{Results}

\subsection{Single Hubbard tripod}\label{sec:Hubbard_tripod}

\begin{figure}[t]
  \centering
  \includegraphics[width=0.9\columnwidth]{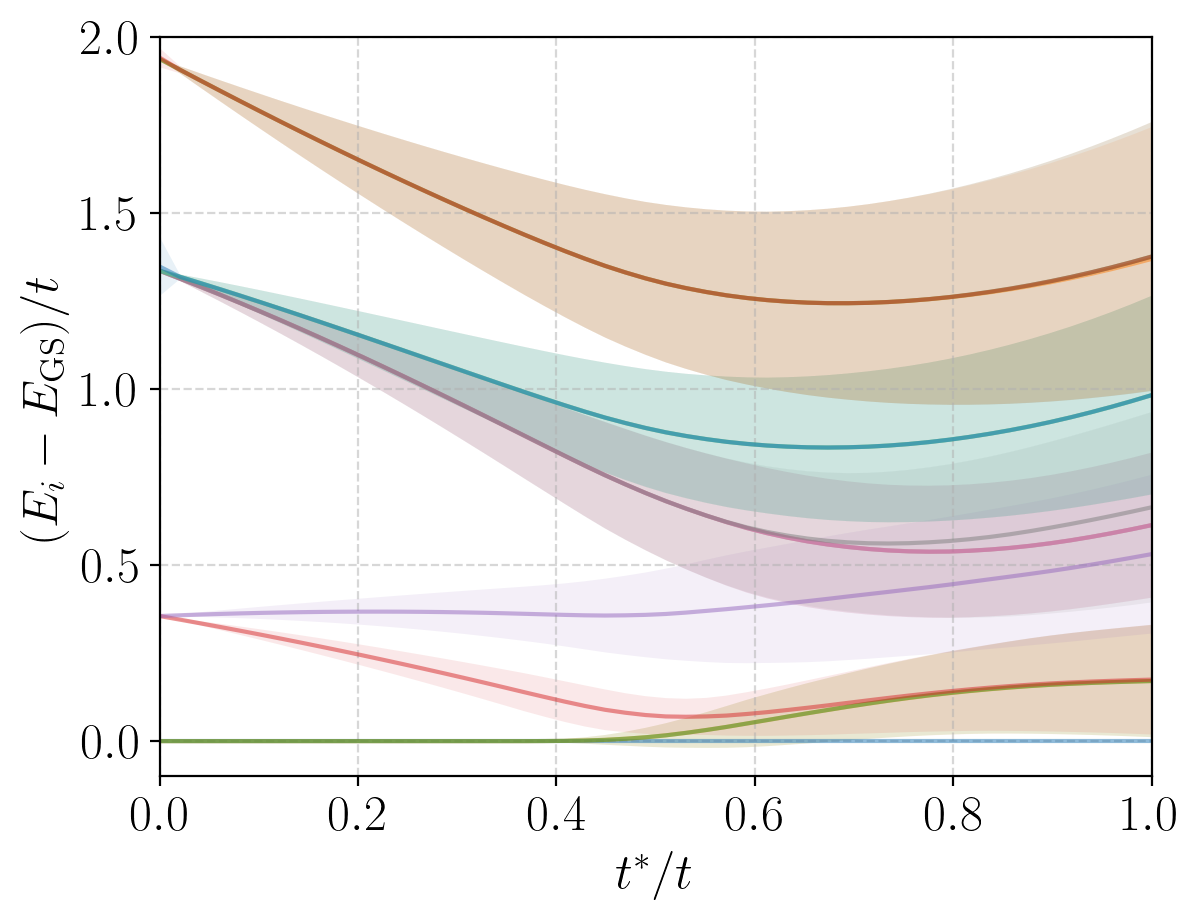}
  \caption{Excitation spectrum of a single tripod subject to random onsite and bond disorder of characteristic strength $t^*$. Solid lines denote the mean values, and shaded areas indicate the distribution's spread at one standard deviation. The data shown represent statistics from 200 individual disorder configurations.}
  \label{fig:tripod_spectrum_dirty}
\end{figure}

A single tripod consists of a central site $c$ connected to three legs $l_1,l_2$, and $l_3$ by hopping $t$, as shown in the left panel of Fig.~\ref{fig:tripod_spectrum}. The tripod Hamiltonian is given by
\begin{equation}
H_0= -t \sum_{\langle i,j\rangle,\sigma}
(c_{i\sigma}^\dagger c_{j\sigma}+\text{h.c.}) + U\sum_i \hat{n}_{i\uparrow}\hat{n}_{i\downarrow},
\label{eq:hubbard_tripod}
\end{equation}
where the sum runs over 
the legs of the tripod, the operator $c^{(\dagger)}_{i\sigma}$ annihilates (creates) an electron with spin $\sigma$, and $n_{i,\sigma}=c^{\dagger}_{i\sigma}c^{\phantom{\dagger}}_{i\sigma}$ is the particle number operator. 
In the simplest case, all hoppings are spin independent and have the same magnitude $t$, which we take as the energy scale throughout the paper. The on-site repulsion $U$ sets the interaction strength. 
The Hubbard model for these configurations has been extensively studied analytically \cite{vanDongen1991_single_star}. By Lieb's theorem \cite{lieb1989}, the Hubbard model on bipartite lattices is characterized by a well-defined 
spin in its ground state at half filling, given by the sublattice imbalance, $S=\frac{1}{2}|N_A - N_B|$. For the particular case of a tripod, this means that the ground state realizes spin $S=1$. As shown in Fig.~\ref{fig:tripod_spectrum}, for any $U>0$, the ground state is well-defined and threefold degenerate. The energy gap to the first excited state is close to maximal at $U=3t$; thus, we consider this value in the remainder of this paper, unless otherwise stated.
Interestingly, the ground state remains degenerate even in the presence of disorder. We consider the perturbation 
\begin{align}
 H_p = -t^*\sum_{ij,\sigma}h_{ij} c_{i\sigma}^\dagger c_{j\sigma}+\text{h.c.},
\end{align}
where $i$ and $j$ run over all pairs of indices, $h_{ij}$ is drawn from the uniform distribution in the interval $[0,1]$ and $t^*$ represents an overall disorder strength. Crucially, this perturbation breaks sublattice symmetry and introduces fluctuations of the on-site potential (from $i\!=\!j$). 
The spectrum of the disordered tripod Hamiltonian $H=H_0+H_p$ plotted in Fig.~\ref{fig:tripod_spectrum_dirty} reveals that the ground state remains nearly degenerate up to a relatively high disorder strength $t^*/t\approx0.5$.

We note that the degeneracy remains exact if one considers spin-orbit coupling on bonds that do not violate sublattice symmetry, that is, on bonds connecting the central site with one of the legs.
In addition, fluctuations in the interaction strength $U$ do not affect the degeneracy of the ground state. By contrast, generic spin-orbit coupling and terms that break time-reversal symmetry, such as an external field, lift the degeneracy of the ground state in the fashion of a Zeeman-splitting, as expected for a spin $S=1$.  
Thus, the robust nature of the ground-state manifold of the Hubbard tripods 
offers an excellent sandbox to explore $S=1$ spin models.

\subsection{$S=1$ Bilinear-biquadratic dimer}\label{sec:two-tripods}

\begin{figure}[t]
  \centering
  \includegraphics[width=0.9\columnwidth]{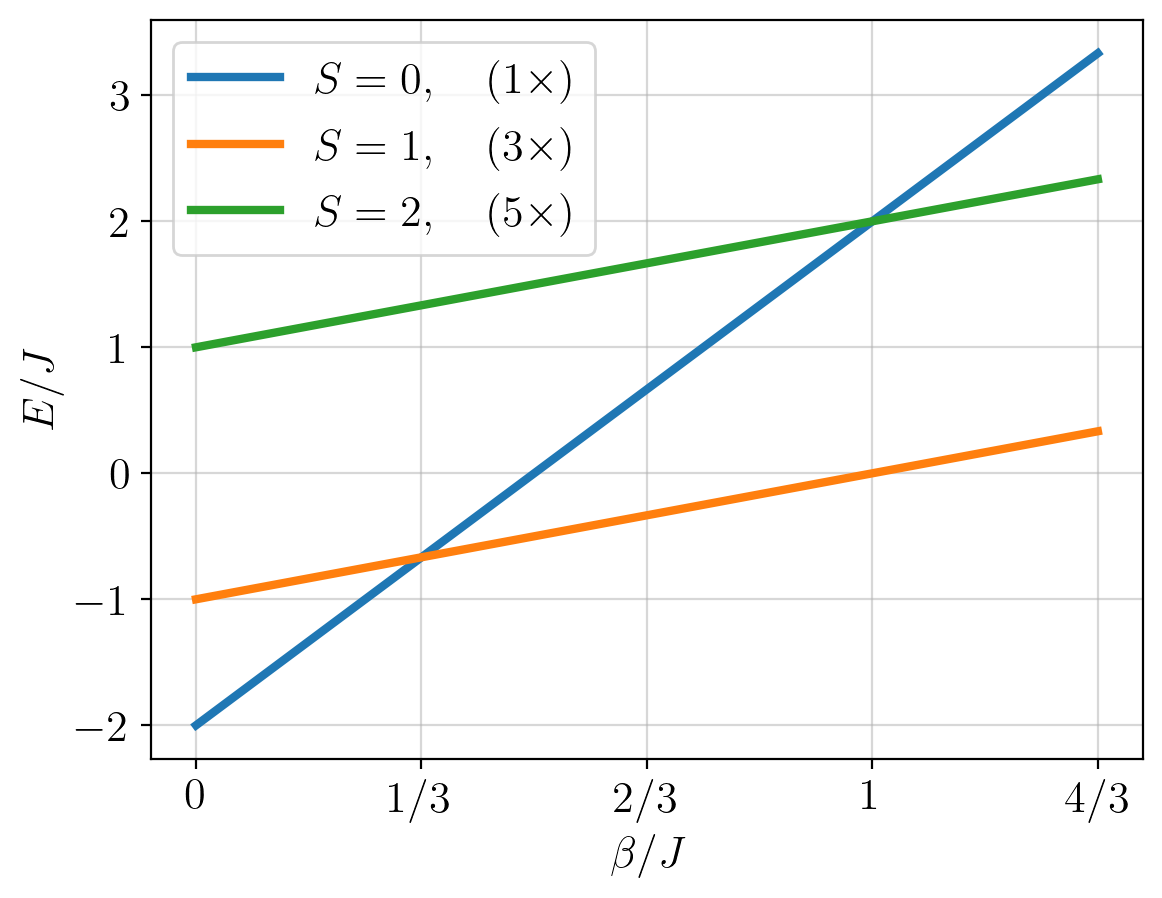}
  \caption{Energy spectrum of the bilinear-biquadratic Hamiltonian of the spin-1 dimer as function of the coupling ratio $\beta/J$. For $J>0$ below (above) $\beta/J=1/3$, the non-degenerate singlet (threefold degenerate triplet) is the ground state. At $\beta/J=1/3$, the singlet and triplet states become degenerate, resulting in a fourfold degenerate ground state separated by a gap of $2J$ from the $S=2$ multiplet. }
  \label{fig:two_spin_spectrum}
\end{figure}

The generic bilinear-biquadratic model
\begin{align}
 \mathcal{H}_{\text{two}} = J\mathbf{S}_1\cdot\mathbf{S}_2+\beta\left(\mathbf{S}_1\cdot\mathbf{S}_2\right)^2,
\label{eq:BLBQ}
\end{align}
with $\mathbf{S}_i$ denoting $S=1$ quantum spin operators on sites $i\in\{1,2\}$, can be solved without difficulty. Since (\ref{eq:BLBQ}) has full spin-rotation symmetry, the eigenstates can be classified according to the irreducible representations of SU(2), and the nine-dimensional representation $D^{(1)}\otimes D^{(1)}$ of the two $S=1$ decomposes as $D^{(0)}\oplus D^{(1)}\oplus D^{(2)}$, dividing the spectrum into a unique singlet ($S=0$), a threefold degenerate triplet ($S=1$), and a fivefold degenerate quintet ($S=2$) manifold. The energies of the different spin-multiplets are $E_{\rm S=0}=-2J + 4 \beta$, $E_{\rm S=1}=-J + \beta$, and $E_{\rm S=2}=J + \beta$. For any antiferromagnetic bilinear coupling ($J>0$), the ground state is either the singlet or the triplet.
Below $\beta/J=1/3$, the singlet has the lowest energy, and above it the triplet. At the special point of  $\beta/J=1/3$, the ground state is the four-fold degenerate mixture of the singlet and triplets with $E_{\rm S=0}=E_{\rm S=1}=-\frac 2 3 J$. We plot the spectrum of the spin-1 dimer in Fig.~\ref{fig:two_spin_spectrum} as a function of $\beta/J$.

These simple analytical considerations will serve as guidelines for identifying the coupling strategies of two Hubbard tripods. In particular, we seek to find coupling configurations, both in the two-tripod and in the tripod chain cases, whose effective low-energy description in terms of $S=1$ quantum spins is given by the AKLT Hamiltonian.

\subsection{Coupling two tripods}

\begin{figure}[t]
  \centering
  \includegraphics[width=0.7\columnwidth]{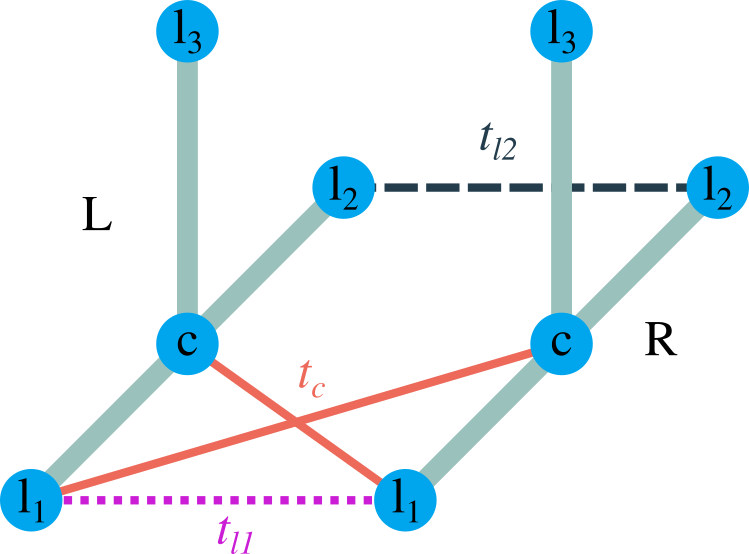}
  \caption{Inter-tripod couplings. Red solid lines depict hopping from a center-site to a leg-site on a different cluster $H_{\text{leg-center}}$, while the dotted magenta line and the dashed dark blue lines denote inequivalent hoppings between legs of different clusters, $H_{\text{leg-leg-1}}$ and $H_{\text{leg-leg-2}}$, respectively.}
  \label{fig:two_tripod_confs}
\end{figure}

We continue with coupling two tripods as depicted in Fig.~\ref{fig:two_tripod_confs}. We label the left and right clusters of sites with $\mathrm{L}$ and $\mathrm{R}$, respectively. 
Within each cluster, each site retains the previously introduced site and spin-specific labels.
Thus, the form of the Hamiltonian of the decoupled system remains the same as (\ref{eq:hubbard_tripod}), but now $\langle i,j\rangle$ denoting pairs that belong to the same tripod.
We consider three distinct hoppings between the tripods, as illustrated in Fig.~\ref{fig:two_tripod_confs}:
\begin{align}
H_{\text{leg-center}} &= -t_c   \sum_\sigma c_{\mathrm{L},c,\sigma}^\dagger c_{\mathrm{R},l_1,\sigma}+c_{\mathrm{R},c,\sigma}^\dagger c_{\mathrm{L},l_1,\sigma}+ \mathrm{h. c.}, \nonumber \\  
H_{\text{leg-leg-1}}  &= -t_{l1}\sum_\sigma c_{\mathrm{L},l_1,\sigma}^\dagger c_{\mathrm{R},l_1,\sigma}+ \mathrm{h. c.}, \\  
H_{\text{leg-leg-2}}  &= -t_{l2}\sum_\sigma c_{\mathrm{L},l_2,\sigma}^\dagger c_{\mathrm{R},l_2,\sigma}+ \mathrm{h. c.}.   \nonumber
\end{align}
The two leg-leg type couplings become inequivalent only when the leg-center coupling is finite. Specifically, the $t_{l1}$ terms connect the same legs that also participate in leg-center hopping, whereas the $t_{l2}$ hopping involves leg-sites that are not coupled to the center of the opposite tripod.

%
Understanding and exploiting this difference is one of the key results of the present manuscript.

\begin{figure}
  \centering
  \includegraphics[width=\columnwidth]{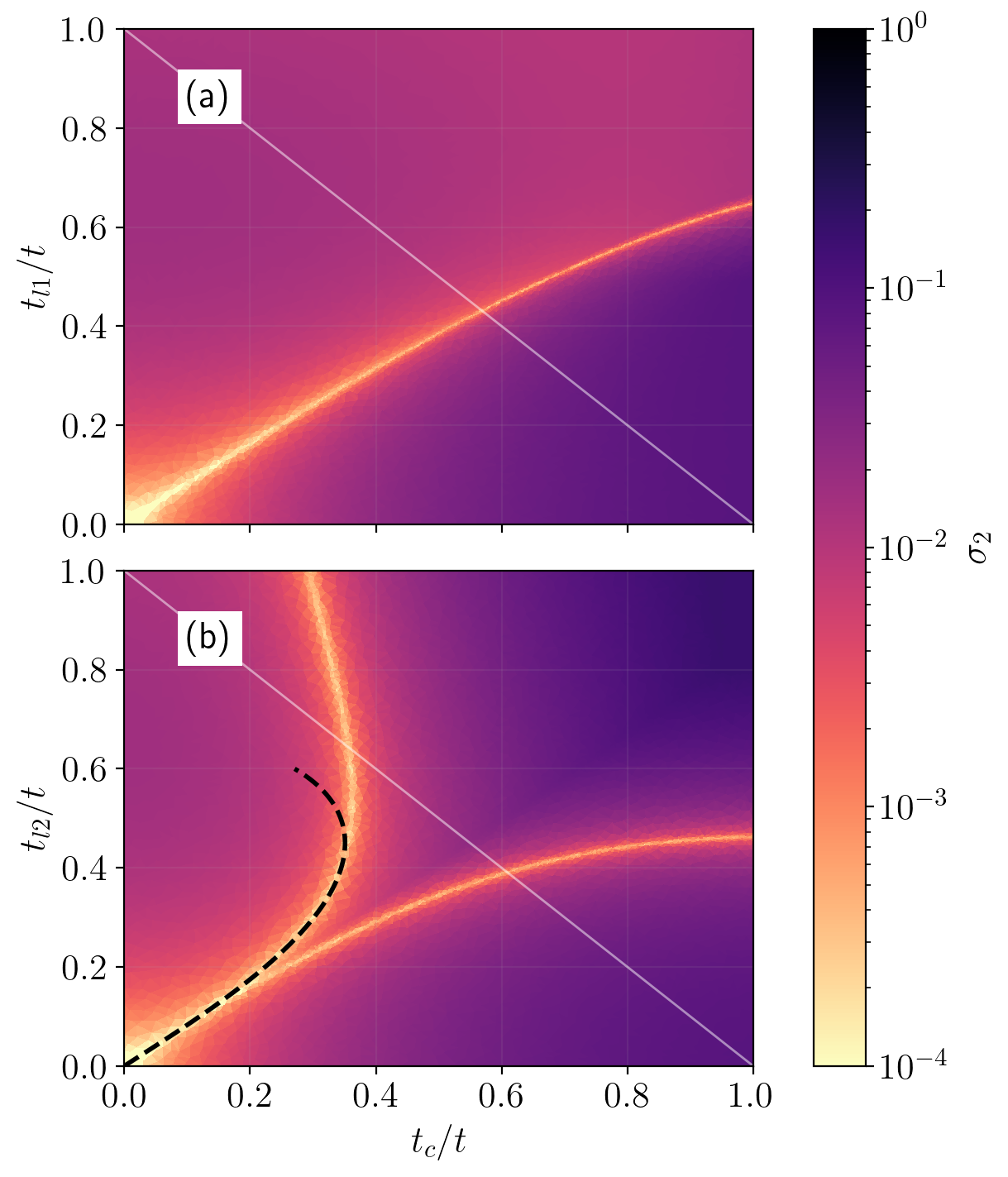}
  \caption{Degeneracy parameter $\sigma_2$ as a function of various coupling strengths. In (a), results are obtained for $t_{l2}=0$, while in (b), $t_{l1}=0$. Diagonal white lines in each plot indicate the sampled section of parameter space in Fig.~\ref{fig:two_tripod_spectra}. The dashed curve in (b) corresponds to results from fourth-order perturbation theory.}
  \label{fig:two_tripod_sigma2}
\end{figure}
\begin{figure}
  \centering
  \includegraphics[width=\columnwidth]{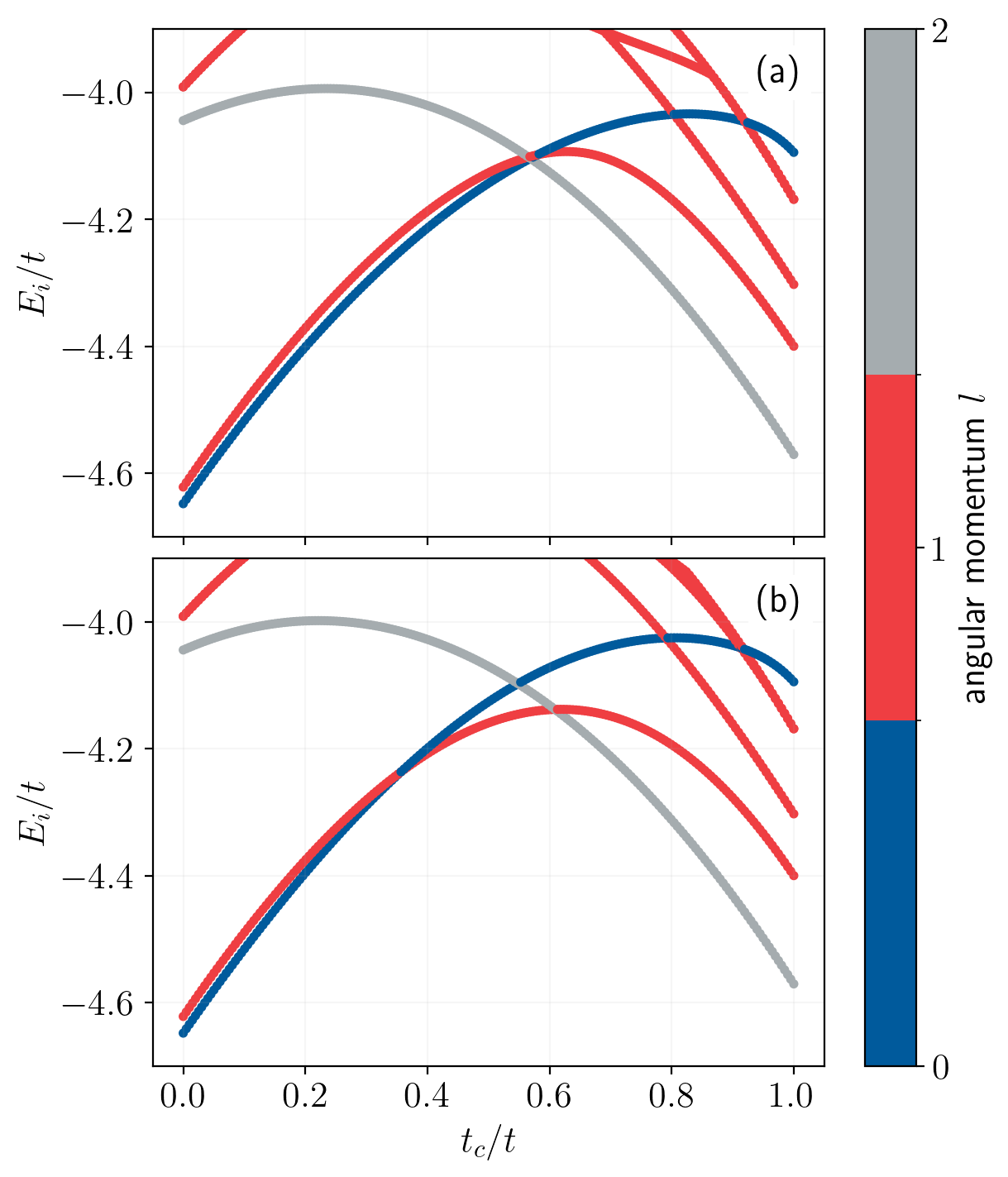}
  \caption{Spectrum of the coupled tripod system along cuts in parameter space. The cut along which the spectrum is calculated is denoted by white diagonal lines on the corresponding subplots of Fig.~\ref{fig:two_tripod_sigma2}. Colours represent angular momentum quantum number $l$, indicative of the $2l+1$ fold degeneracy of a given state.}
  \label{fig:two_tripod_spectra}
\end{figure}

To elucidate the interplay of perturbations, we performed exact diagonalization for clusters at half-filling in the $S_z=0$ subspace. 
This restriction substantially reduces the Hilbert-space dimension while preserving all relevant low-energy sectors, since each $S=0$, $S=1$ and $S=2$  multiplet contains an  $S_z=0$ state.
To track the degeneracy of the ground state in parameter space, we introduce the parameter, $\sigma_2=|E_0-E_1|/2t$, that measures the difference between the two lowest energies.
$\sigma_2$ is vanishingly small around a well-defined curve in the $(t_c\!-\!t_{l1})$-plane, cf. Fig.~\ref{fig:two_tripod_sigma2}(a). Interestingly, when $t_{l2}$ is active in conjunction with $t_c$ as shown in panel (b), the $\sigma_2=0$ manifold splits into two distinct branches from the origin. To better understand the nature of the degenerate $\sigma_2=0$ manifold, we inspect the evolution of the eigenvalues, depicted in Fig.~\ref{fig:two_tripod_spectra}, along the $t_{l1}=-t_c$ and $t_{l2}=-t_c$ cuts in parameter space (denoted by white diagonal lines in Fig.~\ref{fig:two_tripod_sigma2}).

Focusing first on the (a) configuration when $t_{l1}$ is finite, it is evident that around $t_c/t\approx0.6$ all three lowest lying states are degenerate (see Fig.~\ref{fig:two_tripod_spectra}(a)). For smaller values of $t_c$, the ground-state is the singlet, while for large values of $t_c,$ the quintet is the lowest energy state. This means that the aforementioned crossing point separates regions where the effective coupling is antiferromagnetic in the small $t_c$ region and ferromagnetic in the large $t_c$ region. Interestingly, in the limiting case when either $t_c=0$ or $t_{l1}=0$, these conclusions also follow from Lieb's theorem \cite{lieb1989}, as in these cases the coupled clusters have a well-defined sublattice symmetry. 

In the case of finite $t_{l2}$, we find two level crossings. First, at a small $t_c$, the ground state is a singlet. 
Around $t_c/t \approx 0.4$, the singlet and triplet states become degenerate, then after $t_c/t \approx 0.6$, the triplet state gives way to the quintet ground state. That is, in the limiting cases again we have first an effective antiferromagnetic coupling, while large values of $t_c$ gives rise to a ferromagnetic regime. In between, however, is also a region where the triplet state is the ground state. 
As we discussed in the previous section, the ratio $\beta/J=1/3$ in the case of $S=1$ quantum spin doublets leads to a degeneracy between the singlet and triplet states of the joined system. 
Encouragingly, as we observed above, a similar degeneracy can be observed in the case when $t_{l2}$ is varied in conjunction with $t_c$: the upper branch of the $\sigma_2=0$ manifold in Fig.~\ref{fig:two_tripod_sigma2}(b) has precisely this signature.

To further corroborate this statement, we performed quasi-degenerate perturbation theory\cite{day2025} for the Hamiltonian $H=H_{0}+H_{\text{leg-center}}+H_{\text{leg-leg-2}}$, treating $t_c$ and $t_{l2}$ as small parameters. 
This procedure allows us to attain an effective low-energy Hamiltonian in the form of \eqref{eq:BLBQ} acting on the ground-state manifold of the decoupled $H_{0}$. This in turn produces polynomial expressions for the bilinear and biquadratic couplings $J(t_c,t_{l2})$, $\beta(t_c,t_{l2})$. 
We terminate the perturbative series at fourth order, as it is the lowest necessary order needed to produce the biquadratic term. 
Given the $J(t_c,t_{l2})$ and $\beta(t_c,t_{l2})$ polynomials, one can establish a relation between $t_c$ and $t_{l2}$ that guarantees the $\beta/J=1/3$ ratio up to fourth order. We show this curve in Fig.~\ref{fig:two_tripod_sigma2}(b), denoted with dashed lines. The perturbative expression follows the exact diagonalization data closely up to $t_{l2}/t\approx 0.4$. 

\subsection{Forming a chain}

\begin{figure*}[!htbp]
  \centering
  \includegraphics[width=\textwidth]{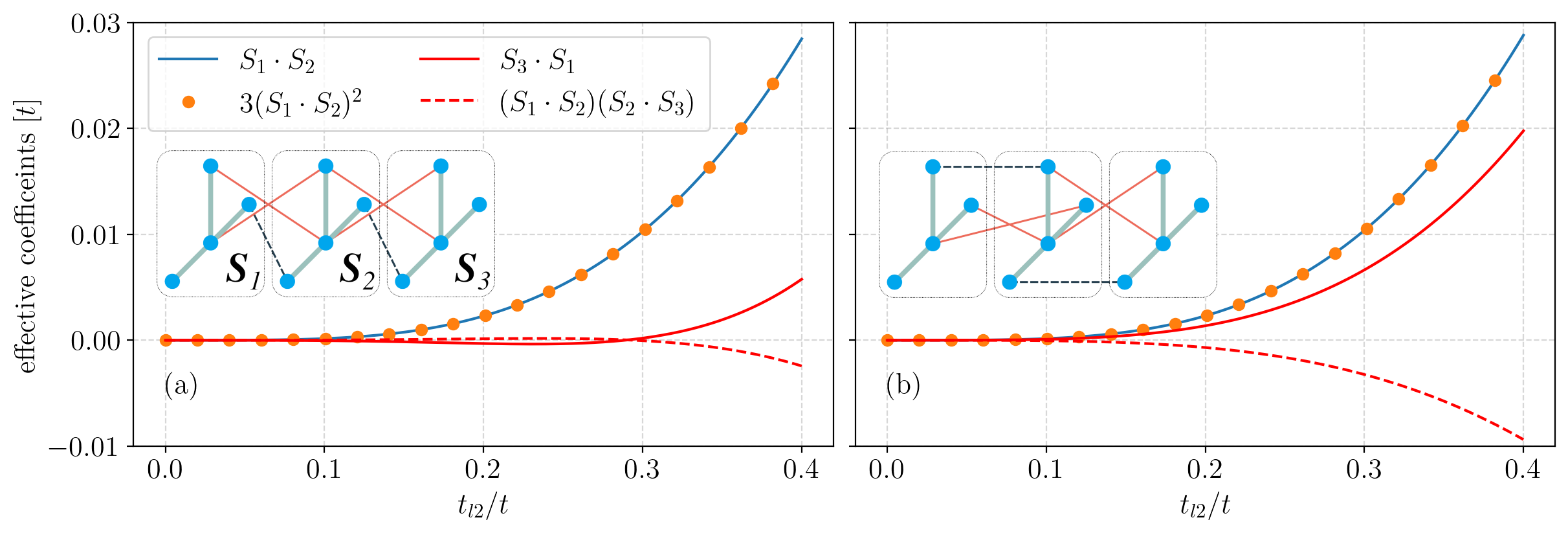}
  \caption{Effective spin model coupling strengths resulting from fourth-order quasi-degenerate perturbation theory for the coupling of three tripods. In (a) and (b), two distinct coupling strategies were considered whose configuration is shown in their respective insets. In the viable coupling scheme of (a), each central site is connected (via $t_c$) to the same leg, and the remaining two legs are coupled in an alternating manner by $t_{2l}$. Whereas, in the unfeasible configuration, each central site couples to two different legs, and contains two inequivalent leg-leg hoppings: one between legs that are not coupled to central sites and one between legs that are also connected to the tripod center.}
  \label{fig:three_tripod_PT}
\end{figure*}

In the previous section, we established that by coupling two tripods through appropriately tuned $t_c$ and $t_{l2}$ hoppings, the low-energy description of the system can be brought to the form necessary for the AKLT model. 
Here we discuss what coupling strategies need to be followed when joining many tripods to form a chain, such that the low-energy Hamiltonian results in \eqref{eq:AKLT_chain}. The heart of the problem we tackle here is whether it is possible to suppress unwanted long-range and multi-spin interaction terms that are, in general, present in the low-energy description. 
To achieve this, we generalized the perturbative procedures described in the previous section and report the numerical data resulting from the fourth-order quasi-degenerate perturbation theory of three coupled tripods. We consider coupling schemes where two neighboring tripods are coupled through $t_c$ and $t_{l2}$ type hoppings, that is, the leg-center and leg-leg bonds that involve different legs. There are five nonequivalent ways one can couple three tripods in such a way. 
In Fig.~\ref{fig:three_tripod_PT}, we show how the low-energy effective spin model parameters behave for two distinct coupling strategies. In both cases, the magnitude of the $t_c$ coupling was tuned so that the ratio of the first nearest neighbor bilinear and biquadratic coupling remains strictly $1/3$, \emph{i.~e.}, moving along the dashed line in Fig.~\ref{fig:two_tripod_sigma2}.
In the viable coupling scheme, illustrated in Fig.~\ref{fig:three_tripod_PT}(a), the central sites are connected to the same type of leg, and the remaining two legs are coupled in an alternating way. For this configuration, the second nearest neighbor bilinear coupling $~\mathbf{S}_1\cdot\mathbf{S}_3$ and the three-site term $~(\mathbf{S}_1\cdot\mathbf{S}_2)(\mathbf{S}_2\cdot\mathbf{S}_3)$ remain suppressed up to $t_{l2}/t\approx0.3$. Distributing the $t_c$ coupling so that each central site connects to two distinct legs results in unwanted couplings (shown in Fig.~\ref{fig:three_tripod_PT}(b)), which are comparable in magnitude to the first nearest neighbor terms. We note that the three further coupling strategies not shown in this figure behave similarly, as can be observed in (b), that is, unwanted terms remain large.
At this point, one might wonder whether more complicated calculations on larger clusters are required before making statements about the infinite-chain limit. In the weak-coupling regime, the perturbative structure itself provides a controlled argument. The low-energy expansion is local and organized as a linked-cluster series in $t_c$ and $t_{l2}$, so the leading coefficients of the unwanted terms are already fixed by the minimal connected cluster where they first appear, namely the three-tripod problem analyzed above. Increasing the chain length does not parametrically enhance these coefficients; it only reproduces the same local contributions on each bond together with higher-order corrections from longer virtual processes. 
Consequently, for the viable coupling strategy of Fig.~\ref{fig:three_tripod_PT}(a), where the leading unwanted couplings are small, these terms are expected to remain subleading even in the thermodynamic limit. Larger-cluster or nonperturbative calculations are still valuable for quantitatively locating the eventual breakdown scale, but they are not necessary for establishing the weak-coupling AKLT regime. Thus, within the perturbative window $t_{l2}/t\lesssim0.3$ and with $t_c$ tuned along the $\beta/J=1/3$ trajectory, the unwanted terms remain subleading in our effective model analysis. 
Based on these observations, we conclude that by extending the coupling strategy shown in Fig.~\ref{fig:three_tripod_PT}(a) into an infinite chain, the low-energy Hamiltonian is well approximated by \eqref{eq:AKLT_chain}. That is, the valence-bond-solid ground state is expected to be realized in the weak-coupling regime of the Hubbard model.

\section{Outlook}
In this work, we established a concrete microscopic route from a fermionic Hubbard model to an effective spin-1 chain with AKLT-type couplings. Starting from a single Hubbard tripod, we identified a robust local $S=1$ manifold and then showed that suitably engineered inter-tripod hoppings can generate the bilinear--biquadratic ratio required by the AKLT point. Extending this analysis to three coupled tripods further demonstrated that specific coupling geometries suppress unwanted longer-range and multispin terms, indicating a viable path toward realizing the valence-bond-solid ground state in an interacting-electron platform.

The next conceptual step is to move beyond one-dimensional AKLT physics toward universal resource states for measurement-based quantum computation. In particular, two-dimensional AKLT-type states with spin $S=3/2$ are known to provide universal MBQC resources \cite{wei2011,wei2012}. This motivates extending the present tripod strategy to tetrapod-like building blocks, where effective $S=3/2$ moments and bilinear, biquadratic, and bicubic interactions are required. 

At the level of two coupled effective spins, the analysis remains feasible for $S=3/2$. When all three inter-cluster hoppings $t_{c}$, $t_{l1}$, and $t_{l2}$ are allowed, they provide enough parametric freedom to obtain the generalized AKLT coupling. A central open challenge is that straightforward perturbative treatments rapidly become prohibitive for increasing cluster size, so complementary techniques will be essential to quantify long-range and multispin corrections in larger arrays.

On the experimental side, highly tunable quantum dot arrays already provide a promising setting for implementing the required hopping architectures \cite{john2024}. Initialization and readout could be integrated through spin-to-charge conversion protocols, building on well-established double-dot and spin-manipulation experiments \cite{ono2002_spin_blockade,petta2005_coherent,koppens2006_driven}. Material platforms with weak and controllable spin-orbit coupling are particularly attractive, including silicon quantum dots \cite{wilamowski2002_byrashba,corna2018_edsr,tanttu2019_soi_control} as well as bilayer-graphene-based devices \cite{banszerus2021_spinvalley,kurzmann2021_kondo}. Together, these developments suggest that Hubbard-engineered AKLT physics is not only theoretically accessible, but also a realistic near-term target for solid-state quantum simulation and, ultimately, resource-state generation for quantum information processing.

\section{Acknowledgments}
C.B. was supported by the Fulbright U.S. Student Program (Study/Research Award, Hungary; Grant No. 2324101).
D.~V. was supported by the Deutsche Forschungsgemeinschaft (DFG, German Research Foundation) under Germany’s Excellence Strategy through the W\"{u}rzburg-Dresden Cluster of Excellence on Complexity and Topology in Quantum Matter – ct.qmat (EXC 2147, project-ids 390858490 and 392019) and the National Research, Development and Innovation Office of Hungary under OTKA grant no. FK 146499. 
D.V. and G. Sz. were supported by the János Bolyai Research Scholarship of the Hungarian Academy of Sciences.
O.~L. was supported by the Ministry of Culture and Innovation and the National Research, Development and Innovation Office within the Quantum Information National Laboratory of Hungary (Grant No. 2022-2.1.1-NL-2022-00004), National Research, Development and Innovation Office (NKFIH) through Grant Nos. K134437 as well as projects KKP133827 and K142179.
This project is supported by the TRILMAX Horizon Europe consortium (Grant No. 101159646).

\bibliographystyle{apsrev4-2}
\bibliography{refs}

\end{document}